# Cognitive Dimensions Analysis of Interfaces for Information Seeking

Gene Golovchinsky
FX Palo Alto Laboratory, Inc.
3400 Hillview Ave, Bldg. 4
Palo Alto, CA 94304  USA

gene@fxpal.com

## ABSTRACT
Cognitive Dimensions is a framework for analyzing human-computer interaction. It is used for meta-analysis, that is, for talking about characteristics of systems without getting bogged down in details of a particular implementation. In this paper, I discuss some of the dimensions of this theory and how they can be applied to analyze information seeking interfaces. The goal of this analysis is to introduce a useful vocabulary that practitioners and researchers can use to describe systems, and to guide interface design toward more usable and useful systems.

## Categories and Subject Descriptors
H.5.2 [**User Interfaces**]: Evaluation/methodology

## General Terms
Human Factors

## 1. INTRODUCTION
Cognitive Dimensions is a technique for analyzing complex information artifacts, including programming languages, device interfaces, and interactive software user interfaces.[3][5] It is designed to be a meta-analysis, a broad-brush approach that looks at structural aspects of the system and identifies characteristics that may impede or enable certain kinds of interactions with the system. It can be used in a summative or formative manner to evaluate existing systems and to drive design of new systems.

In this paper, I apply this tool to the domain of user interfaces for information seeking and exploration. This domain is characterized by complex, cognitively-rich activities. To be effective, information seeking tools need to be designed in a manner consistent with people's cognitive abilities: interfaces that work with people's strengths can be effective even when driven by relatively simple indexing and retrieval schemes; conversely, powerful retrieval engines can be made less usable by coupling them to awkward or ill-designed interfaces.

## 2. SEARCH INTERFACES
Information seeking is an inherently difficult activity due to a number of factors: peoples' information needs are often ill-defined, [1] they may lack the vocabulary required to express the information need, [9] and the need may evolve over time as new information is identified. [6]

These characteristics impose requirements on interfaces through which people look for information. To be useful, interfaces have to be simple to avoid burdening the searcher with distracting or unnecessary complexity, but not too simple to support the cognitive tasks characteristic of information seeking.

## 3. COGNITIVE DIMENSIONS
Cognitive Dimensions is an analytic tool that focuses on the process of interaction, rather than on static analysis of artifacts. [4] It differs from other analytic approaches such as GOMS/KLM in that it does not require a specialized, detailed, and time-consuming analysis that is predicated on very specific interface characteristics. Instead, it allows an interface to be discussed an compared with alternatives using broad terms, represented as different dimensions. The full theory identifies a large number of these dimensions, but only some of them are useful for most analyses of interactive information seeking systems. These dimensions will be discussed below; see [3] for a detailed discussion of all dimensions.

It is important to note that although the dimensions reflect different aspects of interaction, they are not completely orthogonal. In practice, this means that an interface flaw may be reflected simultaneously in more than one dimension.

### 3.1 Premature commitment
This dimension reflects the sequence of steps that a user must perform to achieve a specific outcome. If the user must make a decision early on in some interaction without necessarily having all information to understand the choice, we classify that as premature commitment. For example, being required to provide personally-identifying information prior to being able to interact with a system even in a light-weight manner is an example of premature commitment. So is forcing a user to click on a link in a search result to see some critical piece of information such as the price or an abstract.

Requiring people to ask the system for information that might have just as easily been shown right away was shown to reduce the use of that information. [10] Applied to information retrieval, this suggests that search results should include enough metadata that might help people assess the utility of the document, and accounts for the popularity of snippets as a way of explaining search results. There are limits, of course, to how much information can be presented for each result without making it difficult for the user to understand the information, but designers should consider the tasks that cause people to search, and what information about specific results would make it easier to assess relevance or utility.

## 3.2 Viscosity

Viscosity assesses a design's resistance to change. If, for example, an interface requires the searcher to go through a series of menus or dialog boxes to switch between author search and content search, we say that the design has high viscosity. It is particularly unfortunate if high viscosity is coupled with premature commitment: the user is required to make choices without fully understanding the consequences and it is the difficult to undo these actions once additional information is learned.

Automatic query expansion based on recent browsing history (e.g., [1]) can generate viscosity as the system learns associations between terms and people that may outlast the utility of the association for a particular individual.

Yelp!'s[1] faceted search interface is another example of unnecessary viscosity: A query can be formulated by selecting relevant facets, but once an item is selected, the facet information goes away, forcing the user to backtrack to revise the query. If the original design does not support good viscosity, it may be hard to introduce it later, although there may be measurable benefits to doing it. [7]

## 3.3 Hidden dependencies

This dimension assesses the presence of hidden links among components of the system whose existence may be hard for users to learn. If these links impact people understanding of important system functions, the design should be rated unfavorably on this dimension. For example, "personalized search" learns from users' [8] or groups' [1] interactions with search results about their preferences, and uses that personalization information to affect search rankings. While this approach may improve precision, it may become progressively more difficult to understand why a particular document was or was not retrieved in response to a given query.

This dimension is also related to viscosity. In this case, however, the interface may not reflect to the user that a prior action is now affecting system responses, and it may not be obvious how (or even possible) to undo the effects of hidden dependencies.

## 3.4 Visibility

Visibility reflects how easy it is to view the various aspects of the system. It is related to the notion of affordance. A deep menu system may exhibit poor visibility. For example, Google's Trends search may generate output automatically in response to certain queries (cf. consistency, below) but if the searcher knows that they want to perform a search on structured data, they need to either have to remember the name and search for it, or they need to navigate a deep menu hierarchy (more/even more/labs/Google Trends) to discover the right place to search.

Poor visibility is particularly problematic with faceted search if the user cannot easily add, remove, or refine facet specifications.

## 3.5 Consistency

This is an obvious measure of the degree of similarity of means of accomplishing similar goals in different parts of the interface. It applies to layout (e.g., where people look for the search interface on a web site, where facets selection lists are located, etc.) and to the availability of features. The previous Yelp! example shows a degree of inconsistency because the query refinements are not available on a details page of a search result. Medynskiy et al [7] describe other challenges to making that interface more consistent.

## 3.6 Hard mental operations

Operations that rely on a user's concentrated attention may pose usability problems, particularly when a user may not have the right background knowledge to perform the operation, or may be operating with divided attention. Wolfram|Alpha's minimal interface that requires users to enter syntactically-complex queries is a good example of hard mental operations; Boolean query interfaces (notorious for being error prone for a variety of reasons) are a good example of hard mental operations.

Occurrences of hard mental operations may be exacerbated by high viscosity or premature commitment situations where the user may find it difficult to know what to do or what to undo when an error or unexpected result is observed.

## 3.7 Role-expressiveness

This dimension reflects how well the various visual components of an interface reflect their purpose and the operations available on them. Can the user find the search box? Is it obvious how to compare documents?

While common controls have become reasonably standardized, less common tools such as query expansion or certain kinds of faceted search may require more attention from the interface designer to make the purpose of the controls and the manner in which they should be used obvious, particularly when they are intended to support activity that may involve hard mental operations. The interface should strive for transparency rather than being cluttered with many controls that are used infrequently or whose purpose is not immediately clear.

## 3.8 Progressive evaluation

How easy is it for people to assess what they've discovered, how much progress they've made toward their goal? This dimension becomes particularly important for exploratory search. Interfaces that make it difficult to see which documents have been 'saved' or bookmarked, or ones that hide users' query history, requiring additional interaction to see what has been done may hamper people's exploratory search activities, because being able to get an overview of what has been found or query tactics that have been used may be a useful tool for assessing progress toward satisfying the information need that motivated the search in the first place.

---

[1] www.yelp.com

## 4. CONCLUSIONS

T.R.G. Green's Cognitive Dimensions Theory offers an interesting and powerful toolbox that can be used to characterize and reason about search interfaces without descending into the minutia of particular designs. The vocabulary of cognitive dimensions can form an effective shorthand for expressing complex characteristics of interfaces and systems, and therefore can improve communication between designers, system builders, and other stakeholders. While it was designed for broad applicability to information artifacts of all kinds, it is particularly useful for characterizing the kinds of complex systems that people are using to fulfill their information needs.